\documentclass[12pt]{article}
\usepackage{graphics}
\usepackage{amsfonts}
\usepackage{amsmath}
\usepackage{amssymb}

 \let\b=\beta  \let\g=\gamma   
  \let\h=\eta     
\let\m=\mu    \let\n=\nu         \let\p=\pi    \let\r=\rho
\let\s=\sigma \let\t=\tau

  \let\del=\nabla

\def\\{\hfill\break} \let\==\equiv

\let\0=\noindent

\let\dpr=\partial

\def\qed{\hfill\raise1pt\hbox{\vrule height5pt width5pt depth0pt}}

\def\be{\begin{equation}}
\def\ee{\end{equation}}
\def\bea{\begin{eqnarray}}\def\eea{\end{eqnarray}}

\begin{document}
\markright{Hasen\"ohrl...}

\title{Addendum to ``Hasen\"ohrl and the Equivalence of Mass and Energy"}

\author {
{Tony Rothman\small\it\thanks{tonyrothman@gmail.com }}
\\[2mm]
~ \it NYU, Dept. of Applied Physics, (retired)}

\date{{\small   \LaTeX-ed \today}}

\maketitle

\begin{abstract}
This addendum addresses several objections regarding accelerating cavities in the paper by Boughn and Rothman, ``Hasen\"ohrl and the Equivalence of Mass and Energy," arXiv:1108.2250.

 \vspace*{5mm} \noindent PACS: 03.30.+p, 01.65.+g, 03.50.De

\\ Keywords: Hasen\"ohrl, Poincar\'e, Abraham, Einstein, Fermi, mass-energy equivalence, blackbody radiation, special relativity
\end{abstract}

\setcounter{equation}{0}
\baselineskip 8mm

The paper ``Hasen\"ohrl and the Equivalence of Mass and Energy," by Stephen Boughn and Tony Rothman (BR)\cite{BR} concerned the early attempts by Fritz Hasen\"ohrl to derive the relationship between the energy of black body radiation in a cavity and its equivalent mass.  Section 5.2 of that paper was an attempt to give a relativistically correct derivation for slowly accelerating cavities.  The derivation, however, contained one mistake and also brought criticisms about a further point.  This  note attempts to correct the error and also addresses the second criticism.

 The external force density (force per unit volume) on a radiation-fluid element is given by $\mathfrak  f_\m = T_{\m\n}\, ^{, \n}$, or as in BR (5.14),
\be
\mathfrak  f_\m = \frac{d(\r_o + p_o)}{c^2\, d\t}u_\m + \frac{(\r_o + p_o)}{c^2}\left[\frac{d u_\m}{d\t} + u_\m \frac{\dpr u_\n}{\dpr x_\n}\right] + \frac{\dpr p_o}{\dpr x_\n} \h_{\m\n}.  \label{fm}
\ee
The 4-force density, including heat transfer, can be written as
\be \vec{\mathfrak f} = [(1/c)({\cal F} \cdot {\bf v} + q,_t), \cal F]
\ee
where $\cal F$ is the Newtonian three-force density $q$ is the heat density.  In that case, one gets BR (5.16), the fundamental equation of motion for the radiation:
\be
\mathfrak f_\m = \frac{(\r_o + p_o)}{c^2}\frac{du_\m}{d\t} + \frac{u_\m}{c^2} \frac{dp_o}{d\t}+ \frac{\dpr p_o}{\dpr x_\n}\h_{\m\n} + \frac{\dot q_o u_\m}{c^2}.  \label{fundeq}
\ee

 However, from BR (3.32) the heat term is $\sim E\b^4$, and so when  $ax/c^2 \sim v^2/c^2  < < 1$, which is the approximation assumed in BR, it can be neglected.  Also in that case one can use the equilibrium values of $\r_o$ and $p_o$ (= $\r_o/3$).  For motion in the $x$-direction and retaining terms only of order $a \sim v^2/c^2$,  we have
\be
\mathfrak f' =  \frac{4}{3c^2}\r_o a + \frac{\dpr p_o}{\dpr x}(1 +  \b^2) ,  \label{fundeq2}
\ee
 The total force on the fluid in the lab frame is then:
\be
F= \int_{V'} \mathfrak f' \, dV' =  \int_{V_o} \left[\frac{4}{3c^2}\r_o a + \frac{\dpr p_o}{\dpr x}(1+  \b^2)\right]\g^{-1} dV_o.  \label{fundeq2}
\ee

BR took $\g$ to be the general time-dilation factor
\be
\g = \frac1{(1 + \frac{2ax}{c^2} - \frac{v^2}{c^2})^{1/2}}.
\ee

However, if following  Rindler\cite{R77}, we define $\g$ to be the ratio of the proper cavity length to its  length in the lab frame, there can be only one $\g$, which we take to be $\g^{-1} = (1-\b^2)^{1/2}$. (A "Born rigid" cavity will have  different proper accelerations at each point.  Nevertheless, for such a cavity, boosting to the instantaneous co-moving frame of any fiducial point provides an instantaneous rest frame for the entire cavity, if it is small enough.  Each point in the cavity is undergoing hyperbolic motion, and because at fixed time in the lab frame the different cavity points must be moving with different velocities,  the value of $\g$ might appear to be ambiguous.  However, $\g^{-1}$ at opposite ends differ only at ${\cal O}( \b^4)$, and so which value of $\b$ one chooses is immaterial for this calculation.)  To order $\b^2$ Eq. (\ref{fundeq2}) gives
\be
F= \int_{V_o} \left[\frac{4}{3c^2}\r_o a + \frac{\dpr p_o}{\dpr x}(1+ \frac1{2}\b^2)\right] dV_o,  \label{fundeq3}
\ee
or in terms of $a$
\be
F= \int_{V_o} \left[\frac{4}{3c^2}\r_o a + \frac{\dpr p_o}{\dpr x}(1+ a x)\right] dV_o.  \label{fundeq4}
\ee

The remainder of the derivation goes through  as in BR: To integrate  the second term we use  the divergence theorem $\int_V (\del p_o)\, dV = \oint_A p_o \,\hat {\bf n}\, dA$ for outwardly directed normal $\hat {\bf n}$.  We integrate the third term by parts and get
\be
\int_{V_o}  \frac{\dpr p_o}{\dpr x}(1+  ax ) \, dV_o = \oint_A p_o \hat {\bf n}\, dA + \frac{a}{3c^2} (\r_o V_o)|_{boundary} - \frac{a}{3c^2}\int_{V_o} \r_o \frac{\dpr x}{\dpr x} dV_o.  \label{fundeq5}
\ee
The first term in this equation is the negative of the  external force on the volume element of a fluid. Inserting that term into Eq.(\ref{fundeq4}) and ignoring the last term  gives the relativistic equation of hydrostatic equilibrium when $F = 0$, in agreement with Boughn\cite{Boughn13}. That is,
\be
\oint_A p_o \,\hat {\bf n}\, dA = \frac{4}{3}\frac{E_o}{c^2} a.
\ee
Since the left-hand-side represents a force $= ma$, one immediately has $E_o = 3/4 mc^2$, also in accord with Boughn.
 
However, in this case, if the above assumptions are correct, then the last term in Eq. (\ref{fundeq4}) apparently provides another relativistic correction to the effective mass of the system, lowering the total external force.  That is, if the system is spatially bounded and we extend the integration volume to infinity  where the density and pressure is zero, we can  drop the boundary terms.  Then the total external force  becomes
\be
F = \frac{E_o}{c^2} a,  \label{forcefinal}
\ee
and we immediately have  $E_o = mc^2$,  as desired.\\

Although the neglect of the boundary terms may seem like a risky move, that procedure is employed in the standard proofs of von Laue's and Klein's theorem (see BR \S 4 and Ohanian{\cite{Ohanian12}).  If one takes the point of view that in order for the fluid to be spatially bounded, it must be contained in a physical cavity (which has energy in its own mass density and stresses),  one can proceed as follows.

If the rest energy density of the shell is $T_{00} = \m$,  the stresses are $T_{xx} = \s$ and $T_{0x} = 0$, then the stress-energy tensor for the cavity can be written in  perfect-fluid form\cite{M62}:
\be
(T_{\m\n})_{shell} = (\m + \s)u_{\mu}u_\n + \h_{\m\n}\s
\ee
We then require that the divergence of the combined stress-energy tensor be the force external to the entire system: $f_x= (T_{x\n}\, ^{,\n})_{fluid} + (T_{x\n}\,^{,\n})_{shell}$. Hence, in analogy to above,
\be
F_x = \int_{cav} \left[\frac{\r +p}{c^2} a + \frac{\dpr p}{\dpr x}(1+ax)\right] dV_{cav} + \int_{shell} \left[\frac{\m+\s}{c^2} a + \frac{\dpr \s}{\dpr x}(1+ax)\right] dV_{shell}.
\ee

\begin{figure}[htb]
\vbox{\hfil\scalebox{.6}
{\includegraphics{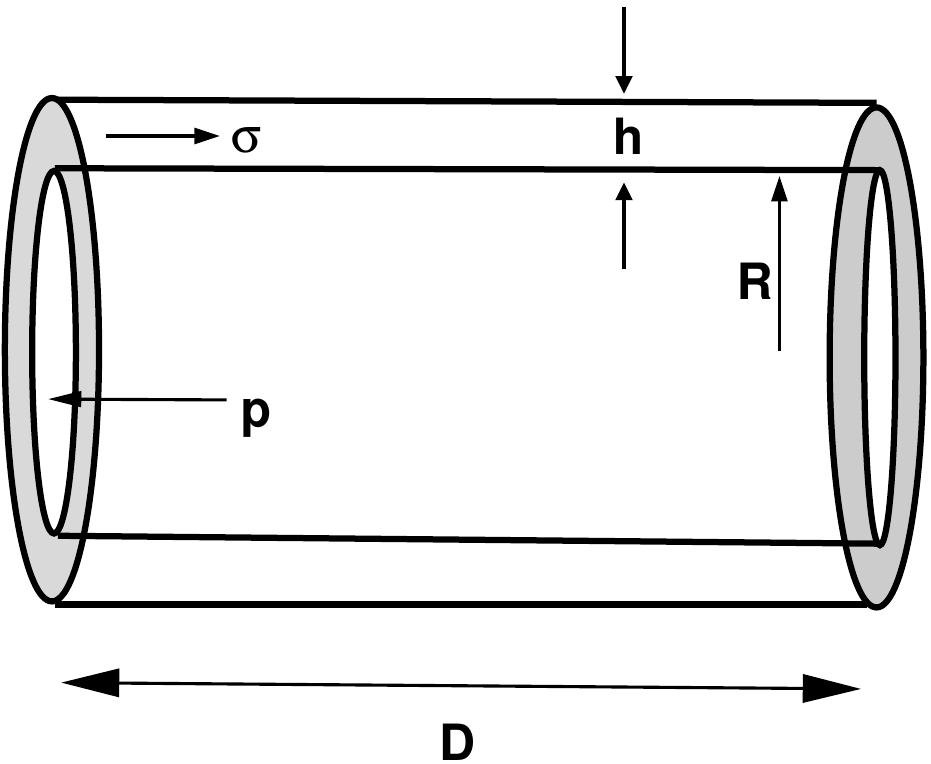}}\hfil}
{\caption{\footnotesize{The fluid is assumed to be held in a cavity of length $D$, radius $R$ and wall thickness $h$.  \label{thickcav}}}}
\end{figure}

Referring to Figure \ref{thickcav}, we see that for the gas to be in contained by the shell, we must have $\p R^2 p = -2\p R h \s$, where $h$ is the thickness of the shell walls and $R$ is the cavity radius.  Thus $\s = -pR/2h$. However, $dV_{cav} = \p R^2 dx$, while $dV_{shell} = 2\p R h dx = (2h/R) dV_{cav}$.  Thus the integrals involving $p$ cancel with those involving $\s$, leaving
\be
F_{ext} = \int_{cav}\frac{\r a}{c^2}dV_{cav}  + \int_{shell}\frac{\m a}{c^2}dV_{shell} = E\frac{a}{c^2}.
\ee
Once again, if $F=ma$, we can immediately conclude that $E=mc^2$ for the entire system.   Consequently, if one ignores the boundary terms in Eq. (\ref{fundeq5}) it apparently becomes  unnecessary to introduce elastic stresses in order to achieve the correct answer for the radiation alone.  In a sense, this appears to be a vindication of Hasen\"ohrl's approach, although because he effectively neglected the relativistic corrections in the fluid's equation of motion, he achieved an incorrect result.\\

{\small{\bf Acknowledgement}\\

Many thanks to Steve Boughn for his continued interest in this problem and and many helpful remarks.

{\small

\end{document}